# Ultrasonic Brain Computer Interfaces for Enhancing Human-Machine Cognition.

**William J. Tyler**
Center for Neuroengineering and Brain Computer Interfaces and Department of Biomedical Engineering, University of Alabama at Birmingham and Heersink UAB School of Medicine, Birmingham, Alabama, USA

## ABSTRACT

Low-intensity transcranial focused ultrasound (tFUS) is rapidly emerging as a transformative non-invasive brain stimulation (NIBS) modality characterized by high spatial resolution and ability to target deep brain circuits. Unlike electromagnetic techniques such as transcranial magnetic stimulation and transcranial direct current stimulation, which are constrained by centimeter-scale resolution and a depth-focality tradeoff, tFUS leverages mechanical pressure waves to modulate both superficial cortical and deep subcortical structures with millimeter precision. This article discusses recent scientific observations and engineering breakthroughs in the advancement of tFUS for next-generation ultrasonic brain-computer interfaces (uBCIs) and human-machine interfaces. These advancements move beyond open-loop systems and demonstrate closed-loop architectures that incorporate real-time electrophysiological feedback to optimize cognitive variables such as attention, learning, trust, and cooperation in various applications. Other advances in the development of ultrasound sensors for sonomyography to decode muscle activation and functional ultrasound to monitor hemodynamic brain activity are discussed as potential elements in bidirectional uBCIs. Together, these advances position ultrasound as a foundational technology for the development of intelligent, adaptive, and bidirectional neural interfaces that will seamlessly integrate human cognition with next-generation automation and robotic systems.

**Running title:** uBCIs for Human-Machine Cognition

## Introduction

The requirement for high-fidelity, non-invasive interfaces with the human central nervous system has driven the exploration and development of ultrasonic neuromodulation [1]. Transcranial focused ultrasound (tFUS) represents a significant advance for noninvasive neuromodulation, offering the capability to focally modulate human deep-brain targets, such as the thalamus [2], amygdala [3], basal ganglia [4], and hippocampus [5] without the risks associated with surgical intervention (**Figure 1A**). While high-intensity focused ultrasound (HIFU) is utilized for thermal ablation in functional neurosurgery, low-intensity tFUS exploits non-thermal mechanical bioeffects to reversibly modulate neuronal excitability [1,6,7]. This precision allows for the causal dissection of human brain circuits and the development of targeted therapies for neurological and psychiatric disorders [1,6-8]. In the context of BCIs and HMIs, tFUS provides a means to "write" information to the brain with unprecedented spatiotemporal specificity, enabling the enhancement of executive functions critical for the operation of semi-autonomous systems in manufacturing, logistics, and transportation [9].

The efficacy of tFUS is governed by the mechanical interaction between ultrasound pressure waves and the heterogeneous, mechanically sensitive biological environment [10]. At the cellular level, ultrasound induced acoustic radiation force (ARF) and shear stress influence neuronal lipid bilayer mechanics. This leads to the activation of mechanosensitive ion channels, like Piezo1 [11], TRAAK [12], TREK [13], and TRP channels [14-16], as well as voltage-gated sodium channels [13,17] and others (**Figure 1B**). These channels transduce mechanical strain into ionic currents at the neuronal membrane [1,10,18]. Other potential explanations include the Bilayer Sonophore Model and Neuronal Intramembrane Cavitation Excitation (NICE) model, which describe how intramembrane space responds to acoustic pressure (**Figure 1B**) [19-21]. Such mechanical fluctuations can induce membrane and ionic channel changes that generate displacement currents sufficient to trigger action potentials [10,17]. Temporal parameters such as pulse repetition frequency (PRF) or duty cycle (DC) also dictate the polarity of neuromodulation outcomes. Studies suggest high DC (e.g., > 50%) or high PRF (e.g., > 500 Hz) are generally associated with excitatory responses and increased spiking rates, whereas low DC (e.g., < 30 %) or low PRF (e.g., < 100 Hz) tend to produce inhibitory effects [22] [4,23-26].

Engineering devices at the scalp interface remains a primary challenge to achieving reliable focal targeting. The human skull is a multi-layered structure, which possesses variable densities across cortical and trabecular bone, that induces significant focal aberration and attenuation of ultrasound. Optimal transcranial tFUS transmission requires the use of fundamental frequencies ($f_0$) below 0.7 MHz. The frequencies most widely used for tFUS are between 250 kHz and 650 kHz, which represents a balance between spatial resolution and transmission efficiency [1]. To mitigate these effects, modern systems utilize computational models combined with multi-element phased arrays for electronic steering and tFUS beam correction (**Figure 1A**) [7,27,28]. Other methods may include the use of subject-specific, 3D-printed acoustic lenses, acoustic metamaterials, or ultrasonic holography to compensate for skull-induced phase shifts [29-32]. System stability for chronic operational can be facilitated by miniaturized transducers, such as capacitive and piezoelectric micromachined ultrasonic transducers (CMUTs/PMUTs), coupled to bioadhesive hydrogels that can maintain stable acoustic coupling for up to 35 days (**Figure 2**) [33,34].

In the periphery, mid-air haptics utilize focused acoustic radiation forces, typically generated by phased arrays of ultrasonic transducers, to non-invasively modulate mechanoreceptors in the skin and induce tactile sensations without physical contact (**Figure 2A**) [35-37]. By providing programmable somatosensory feedback, this technology facilitates advanced human-machine interfaces (HMIs) that reduce visual demand and enhance user agency during gesture-based control of automotive systems, virtual reality environments, and holographic displays [38-40]. While both modalities leverage acoustic energy to alter neural activity, mid-air haptics functions via air-coupled transmission to target the peripheral nervous system, whereas tFUS requires acoustic coupling media (e.g., hydrogels) to efficiently transmit mechanical energy through the cranium for the precise modulation of central brain circuits (**Figure 2**) [1,41-43]. Consequently, the convergence of central neuromodulation via tFUS and peripheral sensory augmentation via mid-air haptics establishes a comprehensive neurotechnological platform for next-generation HMIs designed to optimize human-machine symbiosis and cognitive performance [9,44].

Integrating tFUS capabilities into next-generation ultrasonic brain-computer interfaces (uBCIs) can enable a bidirectional flow of information, effectively linking human cognition with advanced automation. Ultrasound can also serve a unique dual-mode function by modulating, as well as sensing brain and muscle activity. For instance, sonomyography (SMG) can utilize real-time ultrasound imaging to detect muscle deformations for high-fidelity control of robotic systems (**Figure 2B**) [45-47]. Functional ultrasound (fUS) imaging can monitor brain hemodynamics with mesoscopic resolution to "read" complex motor plans [48,49]. These sensing modalities may facilitate the development of unique, closed-loop uBCI architectures described below. Over the past decade, rapid advances in ultrasound neurotechnology, device engineering, and closed-loop sensing architectures have converged to lay the groundwork for uBCIs capable of seamless, high-fidelity integration between human neural circuits and intelligent machines. This review critically examines recent scientific breakthroughs and engineering innovations that are shaping the future of brain-computer interfaces (BCIs), with an emphasis on the transformative potential and remaining challenges of ultrasonic technologies for augmenting human-machine cognition. These intelligent systems may allow for the proactive management of operator states, such as mitigating cognitive fatigue in vehicles or regulating trust during close-proximity human-robot collaborations as described in this perspectives article.

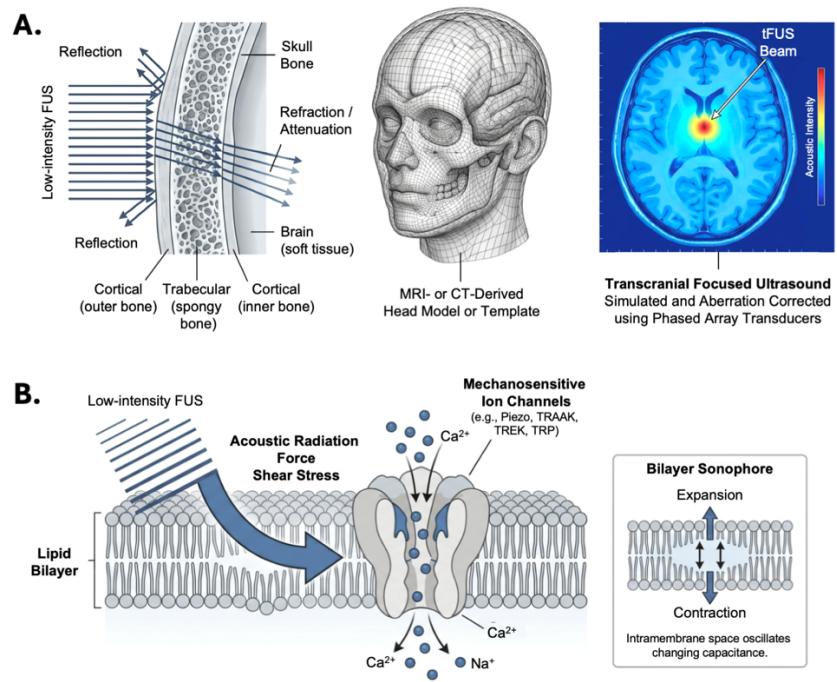

**Figure 1. Biophysical Mechanisms and Targeting of Low-Intensity Transcranial Focused Ultrasound (tFUS). A.** The transmission of low-intensity FUS is significantly impeded by the multi-layered structure of the human skull, which features variable densities across cortical and trabecular bone. These variations induce acoustic reflection, refraction (aberration), and attenuation. To overcome these barriers and achieve millimeter-precision targeting in deep brain structures (e.g., the thalamus), modern systems utilize MRI- or CT-derived head models alongside multi-element phased arrays for electronic beam steering and real-time aberration correction. **B.** At the cellular level, the acoustic pressure waves exert Acoustic Radiation Force (ARF) and shear stress, physically deforming the neuronal lipid bilayer. This mechanical perturbation opens mechanosensitive ion channels (e.g., Piezo1, TRAAK, TREK, and TRP channels), driving an influx of cations like sodium and calcium to alter neuronal excitability. Concurrently, the Bilayer Sonophore and Neuronal Intramembrane Cavitation Excitation (NICE) models suggest that acoustic pressure cycles cause the intramembrane space to rapidly expand and contract, generating capacitance changes and displacement currents sufficient to trigger action potentials without thermal damage.

## Engineering Principles and Safety of Transcranial Focused Ultrasound for Neuromodulation

Understanding the capabilities and limitations of tFUS requires a grasp of the underlying physics of acoustic wave propagation and the engineering solutions developed to control them. This section describes primary technical challenges associated with transcranial transmission and focusing of pulsed ultrasound across the skull to target and modulate the activity of discrete brain regions. The skull has a complex density and unique geometric profile that causes significant attenuation, dispersion, and refraction (focal aberration) of ultrasound waves (**Figure 1A**). Studies have identified the optimal ultrasound frequencies for transcranial transmission and brain absorption are generally at fundamental frequencies ($f_0$) below 0.7 MHz. Ultrasound frequencies for tFUS typically range between 250 kHz and 650 kHz, to balance transmission efficiency with spatial resolution [7,50]. As mentioned, engineering solutions to mitigate aberrations include the use of multi-element phased arrays for electronic steering, the implementation metamaterials and acoustic lenses, and acoustic holography [7,27-32]. While single-element transducers offer simplicity, phased arrays provide the critical capability for correcting phase distortions created by the skull. Acoustic holography represents a further step in complexity but enables the generation of highly specific, arbitrary acoustic fields not achievable with standard array focusing. Irrespective of the beam formation approaches utilized, computational modeling has become an essential component of tFUS planning (**Figure 1A**).

Acoustic simulations, often performed with tools like the k-wave MATLAB toolbox and BabelBrain can be used to predict the pressure field inside the brain before a procedure [51]. Tools like k-Plan and BabelBrain represent sophisticated computational frameworks that integrate subject-specific structural neuroimaging with numerical methods to model acoustic wave propagation and viscoelastic interactions through the heterogeneous cranial environment [28,51,52]. These models rely on high-resolution, subject-specific head models constructed from CT scans or pseudo-CT scans generated from MRI data, which map the acoustic properties (density, speed of sound) of the skull and brain [7,27,53]. These tools are recommended for prospective treatment planning in transcranial focused ultrasound (tFUS) neuromodulation, enabling the calculation of phase corrections to mitigate skull-induced aberrations, the precise estimation of *in situ* focal pressure and thermal deposition for safety compliance, and the optimization of transducer positioning to ensure reliable targeting of deep brain structures [7,28,42]. Other methods can implement acoustic pulses and reflections to characterize the local properties of the skull for implementing real-time phase correction algorithms. Some of these real-time methods can be used to compensate for acoustic attenuation without requiring prior CT or MRI scans, representing a significant step toward making tFUS more accessible and personalized [54].

The temporal characteristics of the acoustic field are defined by the pulse repetition frequency (PRF), which modulates neural circuit dynamics (e.g., 10 Hz to 3000 Hz), and the duty cycle (DC), representing the proportion of active sonication time within a pulse repetition period (Ho et al., 2025; Murphy et al., 2025). Experimental protocols often utilize pulse durations (PD) ranging from microseconds to hundreds of milliseconds, with specific combinations of PRF and DC, such as low DC (~5%) for suppression or higher DC (>30%) for excitation [22,55]. The magnitude of acoustic energy delivered to the brain is rigorously quantified using the spatial-peak pulse-average intensity ($I_{sppa}$) and the spatial-peak temporal-average intensity ($I_{spta}$), which account for the energy within individual pulses and over the total sonication duration, respectively [27]. While "low-intensity" tFUS is generally characterized by intensities below ~ 100 W/cm², human studies typically employ values ranging from approximately 0.5 to 35 W/cm² [7,50,56-58]. Due to the skull's capacity to attenuate acoustic intensity by up to 90%, it is critical to distinguish between free-field parameters (measured in water) and derated or simulated *in situ* values when planning dosing strategies [7,27,42].

To ensure biological safety, protocols are guided by the Mechanical Index (MI), a unitless metric predicting the likelihood of inertial cavitation, defined as the peak rarefactional pressure ($P_r$) divided by the square root of the center frequency [6,27]. The International Transcranial Ultrasonic Stimulation Safety and Standards (ITRUSST) consortium and FDA diagnostic guidelines recommend maintaining an MI ≤ 1.9, an $I_{sppa}$ ≤ 190 W/cm², and an intracranial temperature rise of ≤ 2°C to avoid tissue injury [7,27,58]. One of the primary safety concerns is the potential for skull heating, especially during protocols involving long sonication durations. However, a review of human studies to date shows a strong safety profile. No serious adverse events have been reported in the literature when using low-intensity transcranial focused ultrasound for neuromodulation. Some studies have documented mild and transient symptoms, such as fatigue or headache, which typically resolve shortly after the procedure. To formalize safety standards and reporting conventions ITRUSST continues to work on establishing expert consensus guidelines for the research community. Readers are advised to refer to the recent ITRUSST consensus papers for additional guidelines and safety information [7,27,58].

## Influence of Low-intensity Transcranial Focused Ultrasound on Brain Activity

Recent investigations into tFUS neuromodulation in humans have elucidated parameter-dependent alterations in electrophysiological dynamics and cortical excitability, evidencing both transient modulation and sustained neuroplasticity. High-resolution electroencephalography (EEG) and source imaging have quantified these effects; for instance, tFUS directed at the primary somatosensory cortex (S1; 0.5 MHz, 1 kHz PRF, 36% DC, 23.87 W/cm²) significantly attenuated the amplitude of early somatosensory evoked potentials (SEPs) by attenuating the power of beta and gamma EEG bands [41]. Conversely, excitatory protocols targeting the primary motor cortex (M1) utilizing movement-related cortical potentials (MRCP) demonstrated that lower intensity tFUS at a higher PRF (3 kHz PRF, 5.9 W/cm²) significantly amplified the MRCP source from baseline whereas a lower PRF (300 Hz) did not produce an enhancing effect [22]. Furthermore, theta-burst patterned tFUS (tbTUS; 5 Hz PRF, 10% DC) applied to M1 has been shown to induce long-term potentiation (LTP)-like plasticity, manifesting as sustained increases in motor evoked potential (MEP) amplitudes lasting over 30 minutes, an effect pharmacologically linked to N-methyl-D-aspartate (NMDA) receptor and calcium channel activity [59-61].

Neuroimaging studies utilizing functional magnetic resonance imaging (fMRI) and arterial spin labeling (ASL) have revealed that tFUS modulates large-scale functional connectivity (FC) and perfusion in a target-specific manner, often extending to deep subcortical structures. Kuhn et al. (2023) demonstrated that tFUS (0.65 MHz, 5% DC) targeting the entorhinal cortex (100 Hz PRF) increased regional perfusion by approximately 15.75% and enhanced FC with the dorsolateral prefrontal cortex, whereas amygdala targeting (10 Hz PRF) resulted in localized blood-oxygen-level-dependent (BOLD) signal suppression [62]. Regarding network-level dynamics, tFUS applied to the posterior cingulate cortex (PCC) (0.5 MHz, 10.5 Hz PRF, 5.26% DC) significantly reduced FC along the midline of the default mode network (DMN), correlating with subjective increases in mindfulness [63]. Furthermore, Park et al. (2025) utilized diffusion-weighted MRI (dMRI) alongside resting-state fMRI to quantify structure-function correspondence (SFC) in the DMN following tFUS to the medial prefrontal cortex (mPFC) [24]. They reported that excitatory tFUS (70% DC) significantly enhanced SFC coupling within the DMN, whereas suppressive protocols (5% DC) did not, highlighting the capacity of tFUS to transiently optimize the alignment between structural connectomes and functional dynamics [24].

Biochemical and metabolic analyses have further substantiated the capacity of tFUS to drive offline neuroplastic changes through neurotransmitter modulation. Yaakub et al. (2023) employed magnetic resonance spectroscopy (MRS) to show that theta-burst tFUS (0.5 MHz, 5 Hz PRF, 10% DC, 80 sec duration) delivered to the PCC selectively reduced GABAergic inhibition and increased functional connectivity within the DMN for at least 50 minutes post-stimulation [64]. This reduction in inhibitory tone parallels findings in the thalamus, where tFUS (0.5 MHz, 1 kHz PRF) attenuated the P14 SEP component and modulated beta-band phase distributions [56]. Collectively, these studies establish that tFUS exerts bidirectional control over neural circuits ranging from the suppression of sensory evoked potentials to the induction of LTP-like plasticity and the reorganization of large-scale intrinsic connectivity networks. These effects are governed largely by the temporal characteristics (PRF and DC) of the acoustic energy deposition and brain state or level of task engagement at the time of modulation. Those interested in exploring transcranial ultrasonic neuromodulation are encouraged to consult the literature for protocols and experimental designs that may suit an intended study. As with all neuroscience tools, there are caveats and limitations that should be understood before implementing tFUS methods for research or clinical study [7,27,58].

## Biophysical Mechanisms of Neuromodulation by Focused Ultrasound

The dominant mechanism of tFUS-mediated modulation is hypothesized to be the activation of mechanosensitive ion channels (e.g., Piezo1, TRAAK, TREK) through acoustic radiation force or membrane deformation (**Figure 1B**) [10,11,13,15-18]. The neuronal intramembrane cavitation excitation (NICE) model further suggests that acoustic pressure cycles induce fluctuations in membranes (sonophores) that alter capacitance, generating displacement currents that trigger action potentials (**Figure 1B**) [19,20]. As discussed above, the polarity of the effect (e.g., excitation versus suppression) is highly dependent on temporal parameters such as the PRF, DC, and acoustic intensity profiles [4,22]. High-duty protocols (e.g., 70% DC) are generally associated with excitation, while low-duty protocols (e.g., 5% DC) are often utilized for suppressive effects. While the ability of tFUS to influence neural activity is well-established, the precise biophysical mechanisms are still an active area of investigation. For the low-intensity applications typical of neuromodulation, the effects are understood to be primarily mechanical rather than thermal. In low-intensity tFUS, the induced temperature increase in brain tissue is widely considered negligible, often less than 0.1 °C. Therefore, research to date has focused on non-thermal, mechanical effects of low-intensity ultrasound for neuromodulation.

## Dual Mode Ultrasound for Brain-Computer and Human-Machine Interfaces

The integration of tFUS into BCIs and HMIs can enable a bidirectional flow of information, allowing for real-time cognitive and sensorimotor enhancement [9]. Currently, an array of possibilities exist for the development of ultrasonic brain computer interfaces (uBCIs) that are grounded in scientific observations and engineering feasibility as discussed below. Several studies have shown tFUS can modulate neural substrates involved in cognitive control and motor execution [42,49,65,66]. Targeting the right inferior frontal gyrus (rIFG) with tFUS has been shown to improve behavioral response inhibition, as evidenced by shorter stop-signal reaction times and modulated P300 onset latencies [66]. In the motor domain, theta-burst TUS (tbTUS) can induce long-lasting changes in motor cortex excitability, which could be leveraged to enhance the precision of manual control in industrial teleoperation [42,59,60]. Furthermore, V5-targeted tFUS can enhance feature-based attention to visual motion, significantly reducing error rates in visual-motion-based BCI tasks [67]. These and other observations demonstrate tFUS can target specific brain nodes and networks involved in sensory processing, decision making, action planning, and motor execution that can serve unique functions in different uBCI embodiments.

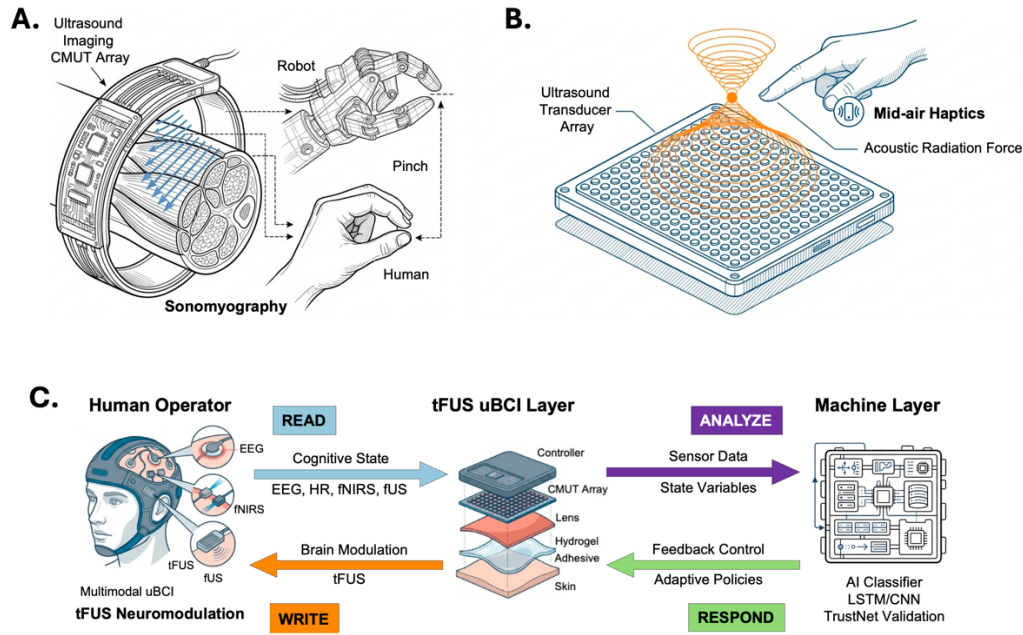

**Figure 2. Peripheral Ultrasonic Neural Interfaces and Closed-Loop Architecture of Ultrasonic Brain-Computer Interfaces (uBCIs). A.** As illustrated, a wearable ultrasound imaging array continuously monitors sub-millimeter morphological deformations in skeletal muscle tissue using a technique known as sonomyography (SMG). By tracking these high-fidelity mechanical signals of spinal motor neuron activation, SMG can accurately decode complex motor patterns, such as pinching and grasping, to control robotic prosthetics. **B.** In other applications of peripheral ultrasound, mid-air haptics can be used to encode tactile sensory experiences. Phased arrays of ultrasound transducers can be programmed to manipulate acoustic wavefronts creating localized pressure nodes of acoustic radiation force in free space. This non-contact form of haptic stimulation triggers mechanoreceptors, rendering complex tactile sensations and virtual buttons that can reduce visual demand in human-machine interactions. **C.** As illustrated, a wearable uBCI seamlessly integrates peripheral and central data streams. The "Read" functions capture cognitive states and motor intent using electrophysiological sensors (e.g., EEG, fNIRS, HR) and functional ultrasound (fUS) imaging. These multimodal data streams are shown being fed into a machine layer that utilizes data classifiers and computational methods to rapidly analyze operator states to instruct the "Write" functions. The system is then shown delivering transcranial focused ultrasound (tFUS) via an encapsulated CMUT array and bioadhesive hydrogel, closing the bidirectional loop between the operator and the machine.

In complex operational environments, such as driving semi-autonomous vehicles or controlling construction equipment, we have proposed that tFUS may be capable of mitigating cognitive fatigue and optimizing situational awareness (**Figures 3A and 4**) [9]. In support of this hypothesis, other noninvasive neuromodulation methods like transcranial random noise stimulation (tRNS) have been applied to the prefrontal cortex to reduce mental fatigue during sustained driving simulations [68]. For teleoperators of drones and robotic systems, tFUS may facilitate rapid learning and decision-making under high psychological stress by modulating arousal and attentional networks [9,66]. Interestingly, EEG-based trust recognition models have been developed to use implicit neural feedback for shaping robot policies in human-robot collaboration (TrustNet), ensuring that autonomous agents adapt to human preferences and cognitive states [69,70]. Combining tFUS with EEG TrustNet frameworks opens the path to unique closed-loop BCI concepts.

A significant advance in ultrasonic neurotechnology is the development of wearable, closed-loop systems that utilize ultrasound for both modulation and biological sensing (**Figure 2**) [32,34]. Closed-loop tFUS systems employ real-time decoding of electrophysiological signals (EEG, LFP, EMG) to trigger on-demand stimulation [71-73]. For example, the NEUSLeeP system integrates a wearable ultrasound patch with polysomnography sensors to deliver targeted stimulation to the subthalamic nucleus (STN) during specific sleep stages, enhancing REM performance [74]. Miniaturized transducers based on capacitive (CMUT) or piezoelectric (PMUT) micromachined technology, coupled with bioadhesive hydrogels, enable stable, long-term neuromodulation outside of clinical settings (**Figure 2C**) [33,34,74,75]. Combining wearable tFUS systems that are comfortable and robust for long-duration applications combined with EEG and other ultrasound sensors begins to enable a new generation of HMIs embodied as uBCIs as described below [71,76,77].

Beyond modulation, ultrasound provides a robust platform for sensing human intent and physiological states [76]. For example, sonomyography (SMG) utilizes ultrasound imaging to detect muscle morphological deformation during contraction, offering higher spatial resolution and deeper muscle monitoring than traditional sEMG [45,76,78,79]. Because muscle fibers are activated by spinal motor neurons, their mechanical motion serves as a high-fidelity mirror of the neural code, allowing for the precise decoding of complex motor plans, such as individual finger movements and multi-degree-of-freedom hand gestures (**Figure 2A**) [76,79]. SMG-based interfaces have demonstrated high accuracy in hand gesture recognition (HGR) for the control of prosthetic limbs and robotic systems, maintaining robustness against muscle fatigue and electrode shift [46,76,79]. In closed-loop uBCI's, SMG can provide one data stream as a "reading" component that informs tFUS "writing" to the motor cortex or subcortical nuclei, such as the subthalamic nucleus (STN), to stabilize motor output or facilitate neurorehabilitation in patients with hemiparesis or spinal cord injury [33,71].

Another peripheral application of ultrasound in HMIs involves non-contact neuromodulation for the encoding of tactile information. Mid-air haptics, particularly through technologies like Ultrahaptics, leverage the acoustic radiation force generated by phased arrays of ultrasonic transducers (**Figure 2B**). These transducers typically operate at fundamental frequencies between 40 and 70 kHz to create noncontact tactile sensations in a three-dimensional volume [39,40,65]. Technically, these systems manipulate ultrasound wavefronts via phase delays to create localized focal points where acoustic pressure induces minute skin tissue displacements, triggering mechanoreceptors [40]. Common implementations utilize 14x14 or 16x16 transducer arrays to generate approximately 10 mN of force, employing modulation frequencies of 200 Hz to align with human vibrotactile sensitivity thresholds [40,43]. Methods such as amplitude modulation, lateral modulation, and spatio-temporal modulation allow for the rendering of complex virtual "buttonscapes" or textured graphics in free space [39,40,65]. Behavioral research in automotive driving simulators has demonstrated that "haptifying" mid-air gestures significantly reduces visual demand—evidenced by a reduction in mean off-road glance time and the frequency of long glances (> 2 sec) while improving secondary task performance through shorter interaction times [40]. Furthermore, recent physiological and neurophysiological studies on noncontact midair ultrasonic stimulation of the median and ulnar nerves have shown the capacity to modulate the autonomic nervous system (ANS), specifically increasing parasympathetic markers like RMSSD [43]. Concurrently, electroencephalography (EEG) data has revealed that such stimulation induces frequency-specific responses in the central autonomic network (CAN), including the insula and dorsal anterior cingulate cortex, as well as significant modulation of default mode network (DMN) hubs like the posterior cingulate cortex [43]. The use of mid-air ultrasonic haptics in HMIs may have unique industrial applications as further described in sections below.

The convergence of ultrasonic "reading" and "writing" technologies is establishing a new paradigm in non-invasive neural engineering, characterized by millimeter-scale spatial precision and the capacity to interface with deep brain circuits [1,49]. Functional ultrasound (fUS) neuroimaging provides a non-invasive means to "read" brain hemodynamics with mesoscopic resolution, enabling the decoding of complex motor plans for uBCI applications [49,67]. Functional US neuroimaging provides a method for "reading" brain activity by monitoring hemodynamic changes in cerebral blood volume (CBV) [48,49,80]. Technically, fUS leverages ultrafast plane-wave imaging (~100 μm spatial and ~1 msec temporal resolution) and singular value decomposition (SVD) spatiotemporal filters to isolate microvascular signals from tissue motion artifacts [48,49,80]. This capability allows for the non-invasive decoding of motor intentions, such as saccadic eye movements and reaching plans, which can then be utilized as trigger signals in closed-loop uBCIs (**Figure 2C**) [49]. Furthermore, photoacoustic tomography (PAT) enhances this sensing suite by combining optical excitation with ultrasonic detection to generate high-contrast biomolecular images, such as hemoglobin mapping and blood perfusion monitoring [44,80]. PAT and fUS together enable a comprehensive "search and rescue" capability for the brain, where dysfunctional circuits are identified through aberrant hemodynamic or metabolic signatures and subsequently rectified via tFUS modulation [80,81]. This integrated approach is particularly effective for modulating feature-based attention and cognitive control, as evidenced by studies targeting visual area V5 to reduce error rates in motion-based BCI tasks [49,67]. Collectively these examples highlight the dual-mode utility of ultrasound to both image and modulate central and peripheral neural activity. These applications and their material characteristics lend themselves to unique embodiments of forthcoming uBCIs as discussed below.

## Potential for Ultrasound in Brain-Computer and Human-Machine Interfaces

The unique capabilities of tFUS including its precision, depth, and parameter-dependent control position it as a transformative technology for creating more powerful and intuitive interfaces between humans and machines (**Figures 3 – 6**). Translating the potential of tFUS into practical, everyday HMI applications requires significant engineering innovation in system design, control architecture, and safety protocols. For chronic, real-world applications outside of a controlled laboratory, tFUS systems must be wearable, comfortable, and unobtrusive. Recent technological advancements are driving this trend. Innovations include soft, flexible, adhesive transducers with integrated acoustic lenses that can conform to the body [33]. At a smaller scale, miniaturized systems based on CMUTs have been developed for use in freely moving animal models, paving the way for similar technology in humans [33,34,74,79,82-84]. Human factors are paramount for wearable devices. A critical component is the coupling medium between the transducer and the skin. To address this, advanced bioadhesive hydrogels and soft conformable designs have been developed that provide stable, long-term acoustic coupling without irritation or discomfort to the user. These tFUS systems, designed for overnight sleep modulation can maintain contact for hours demonstrating practical real-world use of a wearable closed-loop uBCIs that can give rise to other embodiments described [33,74,85].

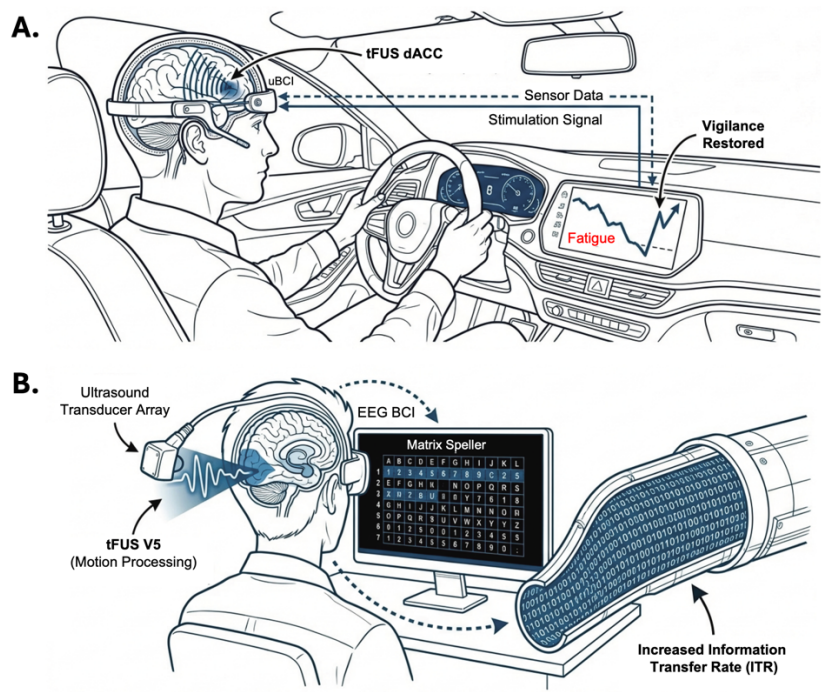

**Figure 3. Applications of Ultrasonic Brain-Computer Interfaces for Vigilance Modulation and Augmentation of Information Transfer. A.** One application to enhance driver safety is the use of uBCIs for dynamic vigilance regulation in semi-autonomous vehicles. As illustrated, a driver supervises a SAE Level 3/4 vehicle while wearing a closed-loop uBCI. The interface continuously monitors physiological indicators (e.g., EEG alpha/theta bands) to detect the onset of cognitive fatigue, mind-wandering, or spontaneous attentional decoupling. Upon detecting a lapse, the uBCI automatically delivers tFUS to brain regions like the dorsal anterior cingulate cortex (dACC) or right inferior frontal gyrus (rIFG) to rapidly restore vigilance, sustain attention, and ensure safe manual handover readiness. **B.** In another application shown, an operator engages with a visual-motion-based matrix speller. As demonstrated by Kosnoff and colleagues (2024) to optimize BCI performance, an ultrasound transducer array was used to deliver tFUS to visual motion processing area (V5). The authors demonstrated tFUS to V5 amplified feature-based attention within the dorsal visual processing pathway and significantly increase the Information Transfer Rate (ITR) during complex BCI spelling tasks [67].

There are numerous potential applications of tFUS in BCIs ranging from directly writing information to the brain to modulating cognitive states critical for human-machine collaboration. Across these applications, tFUS can serve as both a tool to create novel BCI paradigms and a method to enhance existing ones. For example, the potential of tFUS has been realized in the prototyping of a novel non-invasive brain-to-brain interface (BBI). In this case, motor imagery from a "sender" was decoded using electroencephalography (EEG) and used to trigger a tFUS device targeting the primary somatosensory cortex of a "receiver". The tFUS stimulation successfully induced a tactile sensation in the receiver's hand corresponding to the sender's imagined movement, completing a direct communication channel between humans through a BBI [86]. Beyond creating new interfaces, tFUS can also improve the performance of existing BCIs as mentioned. In a study targeting the human visual motion area (V5), applying tFUS was shown to significantly enhance the performance of a visual BCI. This improvement was attributed to the modulation of feature-based attention, demonstrating that tFUS can be used to augment cognitive processes to boost BCI efficacy (**Figure 3B**) [67]. Below several other potential applications are discussed.

## _Enhancing Driver Cognition and Automotive Operation_

EEG can provide a direct, high-temporal-resolution window into a driver's cognitive state by capturing oscillatory brain dynamics and transient event-related potentials (ERPs) [87]. In some embodiments, a uBCI may utilize a "Sense-Analyze-Modulate" closed-loop architecture, where continuous EEG data streams are utilized to detect the neural precursors of cognitive underload, passive fatigue, and spontaneous attentional decoupling (SAD) (**Figure 3A**). Specifically, increases in alpha-band (8–12 Hz) power strongly correlate with attentional withdrawal and the vigilance decrement experienced during monotonous autonomous vehicle supervision [88,89]. Simultaneously, mid-frontal theta (4–8 Hz) activity serves as an index of cognitive control, where decreases in theta power mark mind-wandering and a disengagement from the external driving environment [90]. Furthermore, reductions in the amplitude of the P300 ERP complex reliably indicate diminished attention allocation to infrequent stimuli or hazards [91]. In this uBCI example, continuous EEG features can be fused with peripheral biometrics (e.g., HR/HRV, eye tracking) and fed into machine learning classifiers (e.g., Support Vector Machines) to accurately predict impending cognitive lapses in real-time [92,93].

Upon detecting neural signatures of cognitive underload, a closed-loop uBCI can deliver low-intensity tFUS to the dACC. The dACC is a critical neuroanatomical hub interposed between the DMN and the dorsal attention network (DAN), making it uniquely responsible for conflict monitoring, error correction, and maintaining tonic alertness [94,95]. Causal investigations have revealed that tFUS targeted to the dACC effectively modulates DMN activity, significantly reducing reaction time slowing caused by emotional or environmental distractors, and concurrently elevating parasympathetic markers of heart rate variability HRV [96]. Thus, closed-loop tFUS to the dACC may act as a neurotechnological intervention to rapidly switch the driver from a decoupled, mind-wandering state back to an optimized state of sustained attention and active cognitive control (**Figure 3A**).

Beyond the dACC, several other cortical and subcortical targets can be modulated via tFUS to enhance specific domains of driver cognition and performance. The rIFG is the primary cortical node governing top-down response inhibition, which is the ability to suppress prepotent motor actions, such as failing to brake for a sudden hazard. Experimental evidence demonstrates that tFUS delivered to the rIFG causally shortens stop-signal reaction times (SSRT) by directly modulating the onset latency of the fronto-central P300 ERP complex [66]. Targeting the rIFG is highly efficacious for optimizing driver performance during emergency Takeover Requests (TORs) by accelerating the cognitive switch from passive monitoring to active manual control, thereby reducing resumption lag (**Figure 4**). The insula integrates interoceptive signals and regulates autonomic nervous system activity and sympathetic arousal. Modulating the right anterior insula/frontal operculum (aIns/fO) with tFUS has been shown to reduce parasympathetic fear responses, attenuate emotional distraction interference, and modulate event-locked delta and beta oscillations [96,97]. This target is ideal for calibrating driver arousal to prevent hyper-arousal or panic during complex, high-stress urban driving scenarios. Targeting the primary sensorimotor networks (S1/M1) could enhance attention or the physical execution of driving maneuvers (**Figure 4A**). tFUS directed to S1 has been proven to attenuate SEPs and significantly enhance tactile discrimination, which could optimize human interaction with vehicle steering and hand controls [41]. Furthermore, patterned theta-burst TUS (tbTUS) to M1 induces long-term potentiation-like neuroplasticity that increases motor cortical excitability, which may decrease reaction times for rapid steering or braking adjustments [42,59]. Lastly, V5 is critical for visual motion processing. As described, delivering tFUS to V5 enhances feature-based attention to visual motion by amplifying theta and alpha power within the dorsal visual processing pathway, which significantly improves brain-computer interface performance and reduces error rates during dynamic visual tasks [67]. This could be utilized to improve a driver's visual scanning and ability to track moving hazards in cluttered environments.

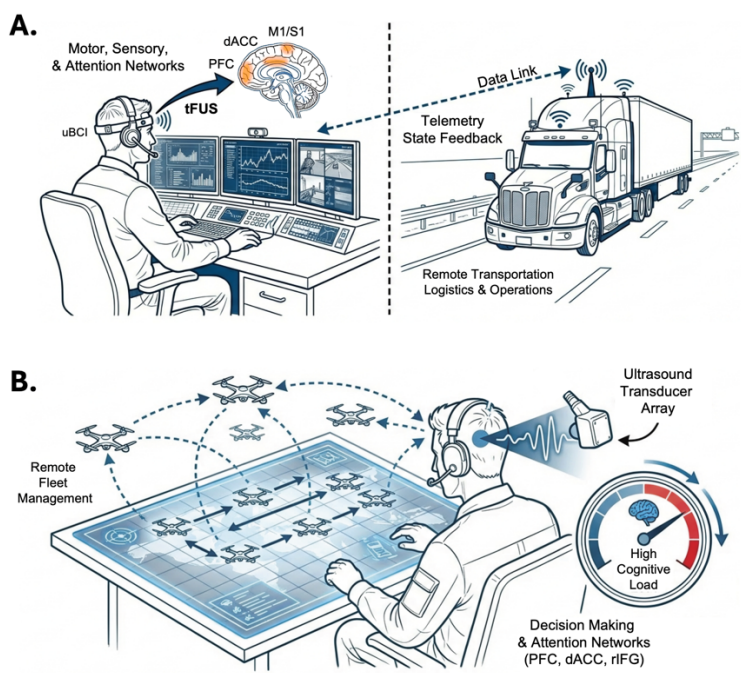

**Figure 4. Neuroergonomic Optimization using Ultrasonic Brain-Computer Interfaces for High-Risk Teleoperation and Multi-Agent Oversight. A.** Another application of uBCIs may be to sustain operator precision in remote logistics. For example, an operator may remotely manage commercial trucking operations via telemetry as illustrated. A wearable uBCI may be used to monitor motor, sensory, and attention networks while delivering tFUS to the primary motor and somatosensory cortices (M1/S1) or the prefrontal cortex (PFC). This precision tFUS modulation may help sustain alertness, fine-tune execution of procedural skills, and shorten reaction times during long-duration, highly demanding teleoperation shifts. **B.** Another application of uBCIs in enhancing human-machine collaboration may be to enhance executive function in during drone management. As illustrated, an operator oversees a remote fleet of drones under a high cognitive load. To mitigate performance degradation, tFUS can be targeted and delivered to the right inferior frontal gyrus (rIFG) or anterior cingulate cortex (ACC) to modulate operator performance. This application may optimize multimodal attention, working memory, and top-down response inhibition, to reduce decision-making errors in multi-agent oversight as depicted.

*Enhancing Human-Machine Collaboration*

The integration of closed-loop uBCIs offers a transformative neuroergonomic approach to mitigating the "automation paradox" inherent in Society of Automotive Engineers (SAE) Level 2 and 3 semi-autonomous vehicles and remote drone fleet operations [9]. Specifically, tFUS targeted to the rIFG causally improves top-down response inhibition by shortening the onset latency of the P300 event-related potential [66]. In applied settings, this neuromodulation can accelerate the "cognitive switch" from passive monitoring to active manual control, significantly reducing resumption lag during emergency TORs [66,98]. Concurrently, targeting the dACC sustains vigilance and conflict monitoring during the monotonous supervision of automated trucking fleets, allowing the uBCI to dynamically recalibrate driver-automation trust and prevent catastrophic states of complacency [93,96]. Furthermore, modulating the PFC, specifically the dorsolateral prefrontal cortex, enhances multimodal attention, working memory, and multitasking throughput, which is essential for remote pilots managing complex logistical operations and swarms of drones where cognitive overload degrades real-time decision-making [99,100].

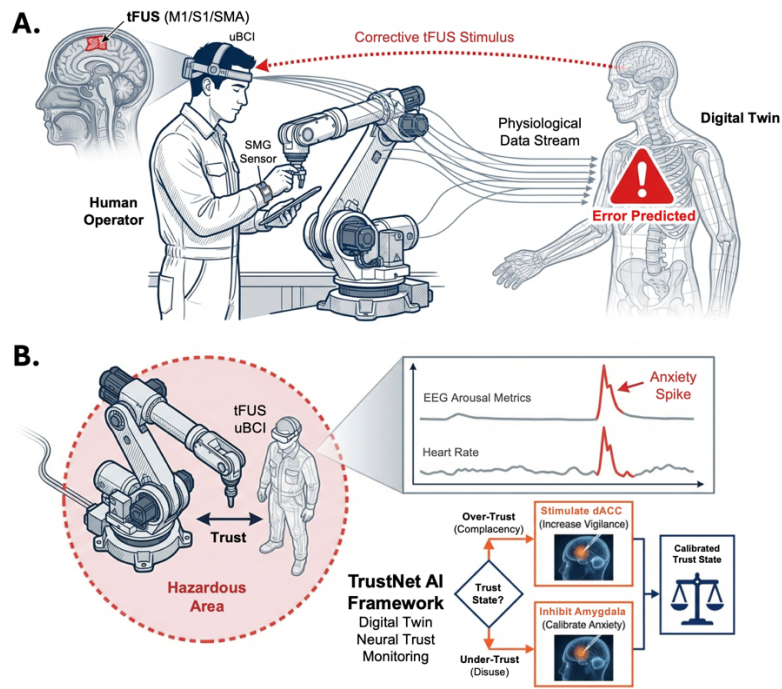

**Figure 5: Cognitive Digital Twin Integration with Ultrasonic Brain-Computer Interfaces for Affective State Regulation in Human-Robot Collaboration. A.** In a manufacturing environment, a human operator is depicted working near a robotic arm. Continuous physiological data (e.g., from EEG and SMG sensors) can be fed to a predictive cognitive digital twin. When the digital twin detects neural or physiological precursors of a performance error, it may trigger a corrective, stabilizing tFUS stimulus targeted to the primary motor cortex (M1) or supplementary motor area to enhance operational safety. **B.** In another expanded application, uBCIs may dynamically regulate trust between humans and machines in industrial environments. For example, a deep ensemble learning algorithm (TrustNet) can evaluate real-time EEG arousal metrics and heart rate/galvanic skin responses to monitor the operator's implicit trust and anxiety levels as illustrated. If dangerous anxiety spikes indicate "under-trust" (potentially causing the operator to disuse robotic automation), a uBCI may deliver suppressive tFUS to the amygdala to offset fear network activation. Conversely, if complacency or "over-trust" is detected, excitatory tFUS stimuli may be targeted and delivered to the dACC to restore cognitive vigilance to calibrate state-dependent human-machine trust.

Beyond executive cognitive control, uBCIs can leverage tFUS to optimize the physical execution of teleoperation maneuvers by targeting the sensorimotor network to support workforce development and human-machine collaboration [9]. Delivering theta-burst tFUS to M1 induces long-term potentiation-like neuroplasticity that increases motor cortical excitability and decreases simple stimulus-response reaction times (Zeng et al., 2022; Bao et al., 2024). Parallel stimulation of S1 significantly attenuates SEPs while paradoxically enhancing tactile discrimination capabilities through local inhibitory surround suppression [41]. For remote operators engaged in mapping, shipping, and fleet management, this precision modulation of M1 and S1 enhances their physical interaction with joysticks, flight controls, and haptic feedback interfaces, refining procedural skill execution and fine motor control (**Figure 5A and 6**) [9,41]. By synthesizing these cognitive and sensorimotor augmentations, tFUS-driven uBCIs provide a comprehensive, neurotechnological framework to ensure resilient and efficient human-machine teaming across the future autonomous transportation workforce [9].

### *Enhancing Trust in Human-Robot Partnerships*

The integration of a "Digital Twin" framework with multimodal biometric sensing, specifically utilizing TrustNet EEG, HR monitoring, and SMG, represents a transformative approach to optimizing human-machine cognition and establishing resilient trust in advanced robotic manufacturing environments (**Figure 5B**). By creating a real-time, personalized computational model of the human operator, the digital twin continuously fuses high-density EEG data processed by "TrustNet," a deep ensemble learning algorithm capable of decoding implicit neural trust states (e.g., over-trust versus under-trust) with accuracies exceeding 88% [69,70,93]. Cardiovascular metrics, such as heart rate and heart rate variability, are synchronously analyzed to index autonomic stress and physiological arousal, providing critical context for the operator's cognitive workload and emotional baseline [101,102]. Concurrently, SMG acts as a high-fidelity biological sensor by employing ultrasound imaging to detect sub-millimeter morphological deformations in skeletal muscle, decoding complex motor intentions while remaining robust against the fatigue and electrode-shift limitations of traditional EMG [76,78]. By synthesizing cognitive trust levels, physiological stress, and physical motor intent, this digital twin framework empowers robotic manufacturing equipment to dynamically adapt its behavioral policies, force-feedback, and operational speeds to match the operator's internal state (**Figure 5A and 6**). This systems-level integration fosters well-calibrated trust, improves industrial safety, and advances workforce development by ensuring human operators and collaborative robots operate in seamless synergy.

To actively manage this human-machine trust dynamic, closed-loop tFUS can be deployed as a highly precise, non-invasive neuromodulatory intervention to correct states of miscalibrated trust (**Figure 5B**). Under-trust, which often leads to the disuse or active neglect of capable robotic systems, is frequently driven by heightened psychophysiological stress, uncertainty, and anxiety. Applying suppressive tFUS protocols to the amygdala has been shown to causally attenuate fear network activation, downregulate hyper-arousal, and reduce subjective anxiety, thereby mitigating the negative affective states that precipitate distrust and facilitating appropriate reliance on the automated system [3,103]. Conversely, over-trust (complacency) poses a severe safety risk characterized by vigilance decrements and cognitive underload, where operators may blindly accept erroneous robotic decisions. To counteract this, targeted tFUS to the dACC or PFC mechanically enhances top-down cognitive control, conflict monitoring, and sustained attention [66,96]. By upregulating these critical executive function hubs, tFUS prevents the degradation of situational awareness, ensuring operators maintain the cognitive vigilance required to critically evaluate robotic actions and execute timely overrides [9]. This integration of precision neurotechnology directly addresses the "ironies of automation," maximizing societal impact by engineering a safer, more adaptable, and highly skilled technological workforce.

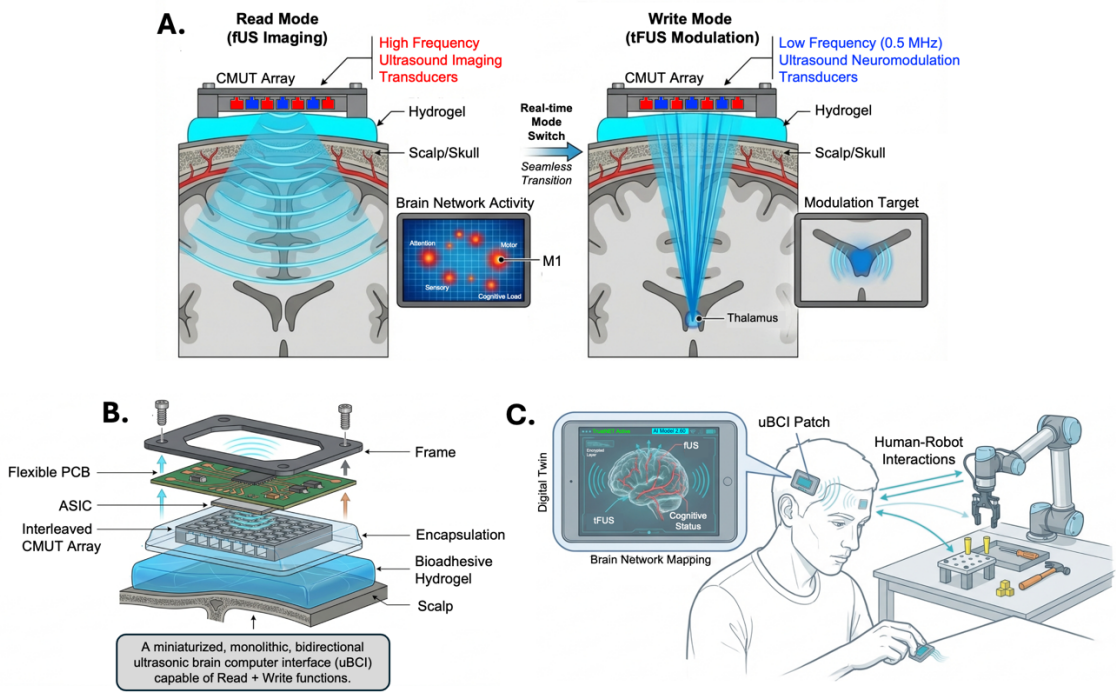

**Figure 6. Architecture of a Bidirectional Ultrasonic Brain-Computer Interface (uBCI). A.** As illustrated, the device may utilize a combination of ultrasound transducers across different frequencies to enable "Read" and "Write" modes. In Read Mode, high-frequency imaging transducers perform functional ultrasound (fUS) imaging to monitor brain network activity and cognitive load. In Write Mode, low-frequency (0.5 MHz) neuromodulation transducers deliver targeted transcranial focused ultrasound (tFUS) to deep brain structures, such as the thalamus. **B.** An exploded view illustrates the hardware stack of the miniaturized uBCI capable of simultaneous Read and Write functions. The layered design may include a structural frame, a flexible printed circuit board (PCB), an application-specific integrated circuit (ASIC), and an interleaved CMUT array. The system may be sealed with encapsulation and coupled to the scalp using a bioadhesive hydrogel to ensure stable long-term acoustic transmission. **C.** The concept for a wearable uBCI patch is shown deployed on a human operator working alongside a robotic arm. Data gathered by the device continuously feeds into a "Cognitive Digital Twin" on a paired interface, dynamically mapping brain networks and monitoring the operator's cognitive status to optimize safety and human-machine collaboration.

## Ethical Considerations on the use of Ultrasonic Brain Computer Interfaces

The ethical design of ultrasonic Brain-Computer Interfaces (uBCIs) for enhancing operator performance requires the systematic integration of the Responsible Research and Innovation (RRI) framework to ensure technological advancements align with fundamental societal values. At the core of this framework is the principle of non-maleficence, which dictates that low-intensity transcranial focused ultrasound (tFUS) parameters must strictly adhere to established safety guidelines—such as FDA limits for spatial-peak temporal-average intensity and IEC mechanical index standards—to prevent thermal or mechanical tissue damage [6,7,27,58]. Furthermore, because uBCIs utilize high-density electroencephalography (EEG) and other continuous biometrics to decode cognitive states, developers must proactively safeguard informational privacy and establish transparent data governance protocols to prevent the unauthorized profiling of an operator's neural data. Ethically designed uBCIs should function as an assistive "intelligent co-pilot" that preserves human agency, dignity, and personal autonomy, ensuring that cognitive augmentation—such as modulating the right inferior frontal gyrus (rIFG) for improved response inhibition—serves to support the operator's intrinsic capabilities rather than overriding their conscious intent [9,66].

In the context of human-machine and human-robot interactions, the ethical use of uBCIs centers on promoting "well-calibrated trust" rather than maximizing human reliance on autonomous systems. Over-trust can lead to complacency and dangerous misuse of semi-autonomous vehicles or manufacturing robots, while under-trust can result in the disuse of capable systems. This ensures that any neurotechnologically induced shift in reliance is grounded in an accurate mental model of the machine's actual reliability, thereby preventing the artificial inflation of trust in a sub-optimal system [69,70,104]. Ultimately, fostering a culture of forward-looking responsibility among neuroengineers, policymakers, and deployers is essential to ensure that uBCIs mitigate the "ironies of automation" equitably, preventing the exacerbation of social inequalities and ensuring that advanced human-machine teaming benefits the broader workforce safely and transparently [93,105].

## Discussion

The emergence of the field of ultrasonic neuromodulation and new applications of ultrasound developed for interfacing with the human nervous system has established tFUS as a transformative, non-invasive technology [1]. Currently, the field is in a highly positive state, characterized by rapid advancements in hardware, software, and diverse applications as discussed. Innovations such as miniaturized CMUTs, bioadhesive hydrogels, and sophisticated closed-loop algorithms have facilitated the development of bidirectional uBCIs to date [33]. These systems uniquely combine precise, deep-brain "writing" capabilities with high-fidelity "reading" modalities, such as SMG for decoding motor intent and fUS for monitoring brain hemodynamics [49,76]. Consequently, uBCIs offer an unprecedented capacity to seamlessly integrate human cognition with next-generation automation and robotic systems, promising to enhance executive functions, mitigate cognitive fatigue, and optimize human-machine symbiosis [9,80].

Despite its remarkable promise, the development and widespread clinical or commercial adoption of uBCIs face several critical limitations and challenges. A primary engineering hurdle remains overcoming acoustic aberration and attenuation induced by the complex density and multi-layered structure of the human skull [28]. While computational models and multi-element phased arrays are currently utilized for beam correction, developing robust, real-time aberration correction algorithms that operate effectively without requiring subject-specific neuroimaging priors is an essential step for broader accessibility [28]. Additionally, while short-term human studies demonstrate a strong safety profile with no serious adverse events, elucidating the long-term bioeffects of chronic tFUS exposure on neural tissue and large-scale network plasticity remains a vital research imperative [103]. Furthermore, establishing comprehensive, application-specific safety standards and reporting conventions, such as those being advanced by the ITRUSST consortium, is urgently required to ensure biological safety and facilitate regulatory approval [7,27,58].

Future directions for uBCI development must directly address these challenges while pushing the boundaries of neurotechnological integration. A critical trajectory involves conducting rigorous longitudinal studies to comprehensively characterize chronic safety and to refine sonication parameters that optimize modulatory effects for diverse cognitive and sensorimotor outcomes [103]. Perhaps the most exciting frontier is the continued miniaturization and engineering of fully integrated, wearable, dual-mode ultrasound interfaces capable of stable, long-term operation outside of controlled laboratory environments [84]. By successfully merging peripheral "reading" sensors with central "writing" modulators, future uBCIs could revolutionize numerous sectors. These advanced closed-loop systems hold the potential to dynamically manage operator states, recalibrate trust in close-proximity human-robot collaborations, and enhance precision in teleoperation, thereby defining the future of resilient, adaptive, and intuitive human-machine interfaces [9,70].

## Conclusions

Transcranial focused ultrasound stands as a potentially transformative technology at the intersection of AI, neuroscience, engineering, technology, and medicine. It offers an unprecedented combination of non-invasiveness, spatial precision, and depth penetration for modulating human brain function. Through continued research into its fundamental mechanisms and refinement of the associated technology, tFUS holds the promise of creating powerful, bidirectional, and intuitive human-machine interfaces. These advanced interfaces could revolutionize applications across society, from enhancing neurorehabilitation in medicine, to improving safety in semi-autonomous transportation, to optimizing performance in industrial collaborative manufacturing.

**Funding**
The author declares that financial support was received for the research and/or publication of this article. This material is based upon work supported by the National Science Foundation under Grant No. NCS-FO-2220677 to WT. The U.S. Government is authorized to reproduce and distribute reprints for Government purposes notwithstanding any copyright notation thereon. The views and conclusions contained herein are those of the author and should not be interpreted as necessarily representing the official policies or endorsements, either expressed or implied, of the U.S. Government.

**Financial Disclosures**
WT is a co-founder of IST, LLC and inventor of patents covering neuromodulation methods and devices, which are described in this manuscript.